\documentclass[amsmath,amssymb]{revtex4}
\usepackage{graphicx}
\begin{document}
\title{Amplification of gravitational waves signal in Michelson coherent-squeezed interferometer}
\author{R. Barak and Y. Ben-Aryeh}
\email{phr65yb@physics.technion.ac.il} \affiliation{Department of
Physics, Technion - Israel Institute of technology, Haifa 32000,
Israel}
\date{\today}
\begin{abstract}
Gravitational waves reaching a Michelson interferometer are expected
to induce a very small change in the length of its arms causing a
phase shift between them, but it is very difficult to observe the
extremely small phase shift signals produced. In the present letter
we show that the gravitational waves signal could be amplified by
orders of magnitude by using very special conditions for a
coherent-squeezed Michelson interferometer in which the coherent
state enters one port of the interferometer and the squeezed vacuum
enters in the other port. We treat the case where without the
gravitational induced phase shift the very strong coherent state
goes out of one output port while the squeezed vacuum goes out the
other output port (the ``dark'' port). While the  phase shift
produced by the gravitation waves does not give any significant
change in the strong coherent output, the light intensity in the
``dark'' port is amplified with decreased fluctuations as the
squeezing increases.
\end{abstract}
\pacs{95.55.Ym,42.50.-p,04.80.-y,07.60.Ly}
\maketitle
A lot of attention has been given in past years to the task of
finding an experimental method, using interferometry, for observing
gravitational waves~\cite{saulson}. Since the effect of terrestrial
gravitational waves on the interferometers' arms length is expected
to be extremely small~\cite{hawking}, gravitational wave detection so far have
not been observed. In the standard use of the Michelson
interferometer a coherent state of light enters one port of the
interferometer and a vacuum state of light enters the other port,
which leads to additional quantum fluctuations in the photon
statistics of the output.

Caves, in his pioneering paper~\cite{caves1,caves2}, was the first to analyze the
properties of Michelson interferometer in which a squeezed vacuum
state replaced the vacuum state entering the interferometer (the
coherent-squeezed interferometer) and showed the advantage of using
such a scheme. Other authors~\cite{kinble,assaf1,assaf2} have shown various properties of
coherent-squeezed interferometers. The amplification of the signal
by the interferometer, usually leads to an amplification of the
noise as well, but this conclusion has been shown to be inaccurate
for phase sensitivity amplification~\cite{jann,zaheer}. In the present letter we
show that under certain conditions the use of a coherent-squeezed
Michelson interferometer leads to an amplification of the
gravitational wave signal by orders of magnitude and at the same
time leads to subpoisson photon statistics.

Let us assume that the state which enters the Michelson
interferometer is given by:
\begin{equation}
|\psi_{\text{in}}\rangle=|\alpha,\zeta\rangle=\hat{D}_1(\alpha)\hat{S}_2(\zeta)|0,0\rangle,
\end{equation}
where~\cite{scully}:
\begin{eqnarray}
\hat{D}_1(\alpha)&=&\exp[\alpha\hat{a}_1^{\dagger}-\alpha^*\hat{a}_1],\\
\hat{S}_2(\zeta)&=&\exp\frac{1}{2}[\zeta^*\hat{a}_2^2-\zeta\hat{a}_2^{\dagger 2}],
\end{eqnarray}
and the subscripts 1 and 2 refer to to the two input ports of the interferometer. 

Neglecting losses in the interferometer the two output operators of the interferometer could be related to the input operators by the unitary transformation:
\begin{eqnarray}
\label{trans}
\left(\begin{array}{c}
        \hat{a}_1 \\
        \hat{a}_2
      \end{array}\right) = \left(\begin{array}{cc}
                      \cos\gamma & \sin\gamma \\
                      -\sin\gamma & \cos\gamma
                    \end{array}\right)  \left(\begin{array}{c}
                                    \hat{b}_1 \\
                                    \hat{b}_2
                                  \end{array}\right),
\end{eqnarray}
where we assumed the special case in which no phase shift is induced in the interferometer. By straight forward algebra we obtain for the output of the coherent-squeezed interferometer:
\begin{equation}
|\psi_{\text{out}}\rangle=\exp(\hat{A}+\hat{B})\hat{D}_1(\beta_1)\hat{D}_2(\beta_2)|0,0\rangle,
\end{equation}
where
\begin{eqnarray}
\beta_1&=&\alpha\cos\gamma;\quad\beta_2=\alpha\sin\gamma,\\
\hat{A}&=&\frac{1}{2}\sin^2\gamma(\zeta^*\hat{b}_1^2-\zeta\hat{b}_1^{\dagger2})+ \frac{1}{2}\cos^2\gamma(\zeta^*\hat{b}_2^2-\zeta\hat{b}_2^{\dagger2}),\\
\hat{B}&=&\sin\gamma\cos\gamma(\zeta\hat{b}_1^{\dagger}\hat{b}_2^{\dagger}-\zeta^*\hat{b}_1\hat{b}_2),
\end{eqnarray}
While the operator $\hat{A}$ represents squeezing effects in the two output ports, the operator $\hat{B}$ represents correlations between the two. These correlations play a fundamental role in the present analysis.

We use the well known Lie-group method~\cite{gilmore1,dasgupta,mufti,zahler1,zahler2} for disentangling the term $\exp(\hat{A}+\hat{B})$ into it's components. We define the operators:
\begin{subequations}
\begin{eqnarray}
\tilde{A}_1&=&\frac{\zeta^*\hat{b}_1^2-\zeta\hat{b}_1^{\dagger2}}{2|\zeta|},\\
\tilde{A}_2&=&\frac{\zeta\hat{b}_2^{\dagger2}-\zeta^*\hat{b}_2^2}{2|\zeta|},\\
\tilde{B}&=&\frac{\zeta\hat{b}_1^{\dagger}\hat{b}_2^{\dagger}-\zeta^*\hat{b}_1\hat{b}_2}{|\zeta|},\\
\tilde{C}&=&\hat{b}_1\hat{b}_2^{\dagger}-\hat{b}_1^{\dagger}\hat{b}_2,
\end{eqnarray}
\end{subequations}
obeying the commutation relations (CR):
\begin{subequations}
\begin{eqnarray}
&&[\tilde{A}_1,\tilde{A}_2]=0,\\
&&[\tilde{A}_1,\tilde{B}]=[\tilde{A}_2,\tilde{B}]=\tilde{C},\\
&&[\tilde{A}_1,\tilde{C}]=[\tilde{A}_2,\tilde{C}]=\tilde{B},\\
&&[\tilde{B},\tilde{C}]=-2(\tilde{A}_1+\tilde{A}_2),
\end{eqnarray}
\end{subequations}
Notice that the CR of the above operators are similar to those of angular momentum operators, where $\tilde{J}_z$ is decomposed into the operators $\tilde{A}_1$ and $\tilde{A}_2$:
\begin{subequations}
\begin{eqnarray}
&&\tilde{A}_1\rightarrow(\hat{I}+2\tilde{J}_z)/2,\\
&&\tilde{A}_2\rightarrow(\hat{I}-2\tilde{J}_z)/2,\\
&&\tilde{B}\rightarrow 2\tilde{J}_x=\tilde{J}_++\tilde{J}_-,\\
&&\tilde{C}\rightarrow 2i\tilde{J}_y=\tilde{J}_+-\tilde{J}_-,
\end{eqnarray}
\end{subequations}

Using the disentanglement method we find after straight forward calculation a certain disentangled form for $|\psi_{\text{out}}\rangle$ given as:
\begin{widetext}
\begin{eqnarray}
\label{exact}|\psi_{\text{out}}\rangle&&=\exp\left\{\eta_-\frac{\tilde{B}-\tilde{C}_2}{2}\right\}
\exp\left\{\eta_+\frac{\tilde{B}+\tilde{C}_2}{2}\right\}
\exp\left\{\eta_2\tilde{A}_2\right\}
\exp\left\{\eta_1\tilde{A}_1\right\}
\hat{D}_1(\beta_1){\rm
\hat{D}}_2(\beta_2)|0,0\rangle,
\end{eqnarray}
\end{widetext}
where
\begin{subequations}
\label{eta}
\begin{eqnarray}
\eta_1&=&\ln\left[{\rm e}^{|\zeta|}\sin^2\gamma+\cos^2\gamma\right],\\
\eta_2&=&\ln\left[\frac{{\rm e}^{|\zeta|}}{{\rm e}^{|\zeta|}\sin^2\gamma+\cos^2\gamma}\right],\\
\eta_-&=&\frac{({\rm
e}^{|\zeta|}-1)\sin\gamma\cos\gamma}{{\rm
e}^{|\zeta|}\sin^2\gamma+\cos^2\gamma},\\
\eta_+&=&\frac{({\rm
e}^{|\zeta|}-1)\sin\gamma\cos\gamma[{\rm
e}^{|\zeta|}\sin^2\gamma+\cos^2\gamma]}{{\rm e}^{|\zeta|}},
\end{eqnarray}
\end{subequations}
We find that according to eq.~(\ref{exact}) the output from the coherent-squeezed Michelson interferometer is given by two squeezed coherent modes with coherent displacement $\beta_1$ and $\beta_2$, and squeezing factors $\eta_1$ and $\eta_2$ for the first and second modes respectively. There also exists an addition of two operators operating on the left of these coherent squeezed modes producing correlations between them. For simplicity of calculation the two operators are $(\tilde{B}+\tilde{C})/2\rightarrow\tilde{J}_+$ and $(\tilde{B}-\tilde{C})/2\rightarrow\tilde{J}_-$ with parameters $\eta_+$ and $\eta_-$ respectively.

In the more common application of gravitational detection using a coherent Michelson interferometer the working point is chosen where all the coherent state exits one port and the vacuum state exits the other port (the ``dark'' port). For the unitary transformation of eq.~(\ref{trans}) such a condition is reached for the case where $\gamma=\pi/2$. In the gravitational wave detection interferometer $|\alpha|$ is a very large number (might be of the order of $10^9-10^{10}$), while the gravitational wave is expected to give a very small phase shift $\delta$ in $\gamma$ (might be of the order of $10^{-9}$ depending on the parameters of the interferometer). Therefor one expects to detect photons in the ``dark'' port as a result of the gravitational phase shift. There are however obstacles~\cite{saulson} to achieve this goal and we show the advantage of the coherent-squeezed interferometer related to the amplification of the gravitational phase shift signal.

Around the point $\gamma=\pi/2+\delta$, where $\delta$ is the phase shift due to the gravitational wave, we can use the first order approximation:
\begin{equation}
\sin(\frac{\pi}{2}+\delta)\approx1;\quad
\cos(\frac{\pi}{2}+\delta)\approx-\delta,
\end{equation}
The coefficients of eq.~(\ref{eta}) for any squeezing value $|\zeta|$ when $\delta\ll1$ are reduced to:
\begin{equation}
\eta_1\approx|\zeta|;\,
\eta_2\approx0;\,
\eta_-\approx({\rm e}^{-|\zeta|}-1)\delta;\,
\eta_+\approx(1-{\rm e}^{|\zeta|})\delta,
\end{equation}
and the operators $\eta_i\tilde{B}$ and $\eta_i\tilde{C}$ ($i=+,-$) in this approximation commute. The output is thus given by:
\begin{widetext}
\begin{equation}
|\psi_{\text{out}}\rangle=\exp\left\{\delta(1-\cosh|\zeta|)\tilde{C}\right\}
\exp\left\{-\delta\sinh|\zeta|\tilde{B}\right\}
\hat{S}_1(\zeta)\hat{D}_1(-\alpha\delta)\hat{D}_2(\alpha)|0,0\rangle,
\end{equation}
\end{widetext}

We find that in the port number 2 of the interferometer the output is the very strong coherent state approximately equal to that of the input state, and so we can refer to it as a classical state such that $\hat{b}_2$ is replaced by $\alpha$ and $\hat{b}_2^{\dagger}$ is replaced by $\alpha^*$. Thus we can look at the ``dark'' port of the Michelson interferometer as a one mode system with $\zeta=r{\rm e}^{i\theta}$ and $\alpha=p{\rm e}^{i\phi}$
\begin{eqnarray}
|\psi_{\text{out}}\rangle_1&=&\hat{D}(\alpha\delta(1-\cosh r))
\hat{D}(-\alpha^*\delta{\rm e}^{i\theta}\sinh r)\times\nonumber\\
&&\hat{S}(\zeta)\hat{D}(-\alpha\delta)|0\rangle_1,
\end{eqnarray}

We rearrange the expression for $|\psi_{\text{out}}\rangle_1$ by first adding the two left terms using the relation~\cite{glauber}:
\begin{equation}
\label{DD}\hat{D}(\alpha_2)\hat{D}(\alpha_1)=\hat{D}(\alpha_1+\alpha_2)\exp\left\{\frac{1}{2}(\alpha_2\alpha_1^*-
\alpha_2^*\alpha_1)\right\},
\end{equation}
then rearranging the new terms using the relation~\cite{fisher}:
\begin{equation}
\hat{D}(\alpha)\hat{S}(\zeta)=\hat{S}(\zeta)\hat{D}(\alpha\cosh
r+\alpha^*{\rm e}^{i\theta}\sinh r),
\end{equation}
and then adding again the two left terms using eq.~(\ref{DD}). After straight forward calculations we get the final result of:
\begin{equation}
\label{out1}|\psi_{\text{out}}\rangle_1=G_1G_2\hat{S}(\zeta){D}(\kappa)|0\rangle_1
\end{equation}
where $G_1$ and $G_2$ are phase terms, which are not important for photon number statistics and are given by:
\begin{subequations}
\begin{eqnarray}
G_1&=&\exp\left\{ip\delta\sinh r\sin(\theta-2\phi)(1-\cosh
r)\right\},\\
G_2&=&\exp\left\{i|\alpha|^2\delta^2\sinh
r\left[\sin(\theta-2\phi)\right.\right.\nonumber\\
&&-\left.\left.2\cosh r\sin(\theta-2\phi)\right]\right\},
\end{eqnarray}
\end{subequations}
and the coherent parameter is given by:
\begin{widetext}
\begin{equation}
\kappa=\alpha\delta\left[\left(\cosh
r-\cosh^2 r+{\rm e}^{i(\theta-2\phi)}\sinh r -\sinh^2 r-2{\rm
e}^{i(\theta-2\phi)} \sinh r\cosh r-1\right) \right],
\end{equation}
\end{widetext}
Thus, the output of the ``dark'' port under of the coherent-squeezed Michelson interferometer, under the above conditions, is given by a coherent state with a coherent parameter $\kappa$ (replacing the coherent parameter $-\alpha\delta$ from a coherent Michelson interferometer) squeezed from the left by the original squeezing parameter $\zeta$. For a coherent Michelson interferometer $\zeta=1$ and $\kappa$ is reduced to $\kappa=-\alpha\delta$.

For the special and more common case of a coherent state Michelson interferometer $|\psi_\text{out}\rangle_1$ is given by eq.~(\ref{out1}) with $r=0$ which gives $G_1=G_2=1,\, S(\zeta)=1,\, \kappa=-\alpha\delta$. The output of the ``dark'' port is therefor given by a coherent state with the coherent parameter $-\alpha\delta$.

The point of interest in this paper is the coherent-squeezed michelson interferometer. This case has several interesting results as can be seen from Fig.~\ref{alpha}:
\begin{figure}[htbp]
\includegraphics[width=0.5\textwidth]{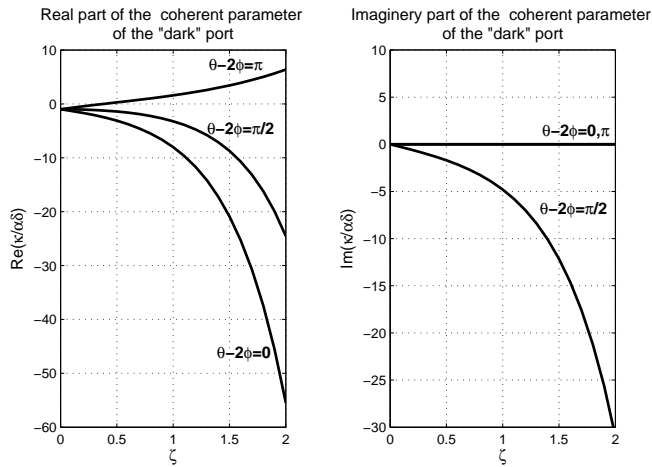}
\caption{\label{alpha}The strength (real and imaginery parts) of the coherent
parameter of the dark port of the interferometer. The maximum is
observed for $\theta=2\phi$.}
\end{figure}
The coherent part of the output is amplified exponentially
depending on the amount of squeezing. This result is critically
dependent on the value of $\theta-2\phi$, reaching a maximum at
$\theta=2\phi$. Other values have both a smaller value for the coherent part and also add a complex part to it.

\begin{figure}[htbp]
\includegraphics[width=0.5\textwidth]{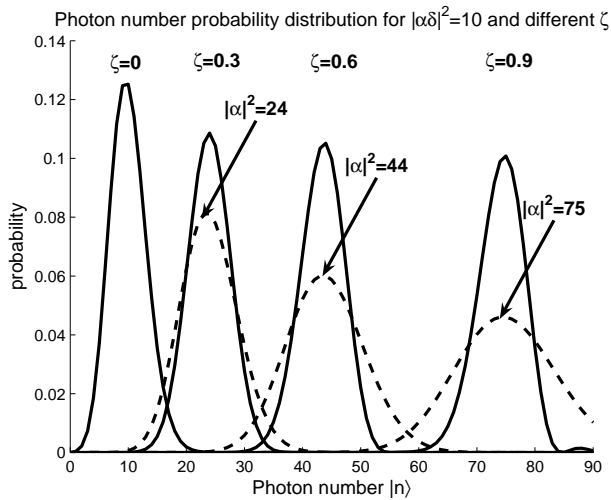}
\caption{\label{subpois}The photon number distribution in the ``dark'' port for
$|\alpha\delta|^2=10$ and different values of $\zeta$
of the squeezed state and $\theta=2\phi$ in the input ports. An
increasing subpoisson distribution is achieved as $\zeta$ is
increased and the maximum point of the photon number distribution is shifted to a higher value of $|n\rangle$, as can be seen from the comparison of the graph for $\zeta=0.3,0.6,0.9$ with the equivalent poisson distribution of the
dashed graph of a coherent state with $|\alpha|^2=24,44,75$.}
\end{figure}

The coherent state is squeezed by the original squeezing parameter.
Under the condition $\theta=2\phi$ we find subpoisson statistics for
large values of the coherent state amplitude and the distribution
becomes more subpoissonian with increased values of squeezing as
shown in figure~\ref{subpois}.
Notice that the maximum of the photon number distribution is found at a point proportional to $\text{Re}(\kappa/\alpha\delta)$ and not to its' square. Although $|\kappa|$ increases exponentially (see Fig.~\ref{alpha}) the maximum photon probability distribution of a squeezed coherent state decreases relative to that of a coherent state with the same coherent parameter (as shown in~\cite{scully}, Fig.3.3). Thus the outcome is a slower increasing photon number subpoisson distribution.

The mean of the distribution is easily calculated:
\begin{equation}
\langle\hat{n}_1\rangle=|\kappa|^2[\cosh 2r-\cos(\theta-2\phi)\sinh 2r]+\sinh ^2r
\end{equation}
Thus for a strong coherent state inserted  in one input port and a squeezed vacuum state inserted in the other input port and under the condition $\theta=2\phi$  the maximum photon probability is found at $|\kappa|^2{\rm e}^{-2r}$ . In the same fashion the standard deviation of the distribution is calculated~\cite{scully}:
\begin{equation}
\langle\triangle\hat{n}_1^2\rangle=|\kappa|^2[\cosh 4r-\cos(\theta-2\phi)\sinh 4r]+2\sinh ^2r\cosh ^2r
\end{equation}
For a strong coherent state inserted in one input port and a squeezed vacuum inserted in the other input port and under the condition 
$\theta=2\phi$  the standard deviation of the distribution is $|\kappa|{\rm e}^{-2r}$  which forms the subpoisson distribution. The subpoisson properties obtained here are similar to those for squeezed coherent states but notice that kappa which replaces alpha  increases as function of squeezing which leads to amplification of the gravitational waves signal.

We have shown in this letter a rigourously exact calculation using  disentanglement theory for the coherent-squeezed Michelson interferometer. Using this calculation we have found the photon number distribution of the ``dark'' port of the interferometer. This distribution is a squeezed coherent state with the same squeezing as the input state and an exponential increasing coherent parameter with the squeezing. The maximum of the photon number distribution also increases with the squeezing and becomes more subpoisson.

\bibliography{ar4}

\end{document}